# The STARFLAG handbook on collective animal behaviour: Part II, three-dimensional analysis

ANDREA CAVAGNA[1,2], IRENE GIARDINA[1,2], ALBERTO ORLANDI[1], GIORGIO PARISI[1,3], ANDREA PROCACCINI[1,3]

[1] *Centre for Statistical Mechanics and Complexity (SMC), CNR-INFM*
[2] *Istituto dei Sistemi Complessi (ISC), CNR*
[3] *Dipartimento di Fisica, Universita' di Roma 'La Sapienza'*

Correspondence to Andrea Cavagna and Irene Giardina:
SMC, CNR-INFM, Dipartimento di Fisica, Universita' di Roma 'La Sapienza', Piazzale Aldo Moro 2, 00185 Roma, Italy
*andrea.cavagna@roma1.infn.it*
*irene.giardina@roma1.infn.it*

Alberto Orlandi:
SMC, CNR-INFM, Dipartimento di Fisica, Universita' di Roma 'La Sapienza', Piazzale Aldo Moro 2, 00185 Roma, Italy

Giorgio Parisi and Andrea Procaccini:
Dipartimento di Fisica, Universita' di Roma 'La Sapienza', Piazzale Aldo Moro 2, 00185 Roma, Italy

## Abstract

The study of collective animal behaviour must progress through a comparison between the theoretical predictions of numerical models and data coming from empirical observations. To this aim it is important to develop methods of three-dimensional (3D) analysis that are at the same time informative about the structure of the group and suitable to empirical data. In fact, empirical data are considerably noisier than numerical data, and they are subject to several constraints. We review here the tools of analysis used by the STARFLAG project to characterize the 3D structure of large flocks of starlings in the field. We show how to avoid the most common pitfalls in the quantitative analysis of 3D animal groups, with particular attention to the problem of the bias introduced by the border of the group. By means of practical examples, we demonstrate that neglecting border effects gives rise to artefacts when studying the 3D structure of a group. Moreover, we show that mathematical rigour is essential to distinguish important biological properties from trivial geometric features of animal groups.

Collective animal behaviour is a fascinating phenomenon that has interested scientists for a long time (see, for a review, Krause & Ruxton 2002). How does global co-ordination emerge in a flock of thousands birds wheeling over the roost? How do fish in a school organize in a milling pattern? Are these self-organized phenomena, or is there a centralized control mechanism? What is the evolutionary function of collective behaviour? What are the common features in different species? What is the precise nature of the interaction among the individuals?

The best way to tackle these questions is through a feedback between theoretical models and empirical observations. However, obtaining three-dimensional (3D) data about large groups of animals in the field has been very difficult up to now, and for this reason the vast majority of studies focused on 3D numerical models. There has been little space for a comparison between experiments and theory. In fact, comparisons have mostly been made between different kinds of models (Parrish et al. 2002), rather than between models and empirical data. The consequence of this fact is that most of the quantitative tools normally used to characterize collective animal behaviour are tailored to numerical models, and in some cases are unsuitable to 3D empirical data sets. Many of the parameters that are varied when studying a numerical model, cannot be changed during empirical observations. The typical example is the amount of noise each individual is subject to when deciding its direction of motion: this is just an easy knob to turn in a numerical model, but it is definitely not possible to do so for empirical observations. Even those quantities that do change in the field, as group size, velocity, nearest neighbour distance and so on, unfortunately do not change on the observer's demand. Moreover, empirical data are necessarily noisier than numerical data, and this may force us to discard some quantities that are on the other hand useful when studying models.

Recently, the STARFLAG project developed some new techniques able to reconstruct the 3D positions of individual birds in large flocks of starlings in the field (Cavagna et al. 2008). The empirical methods introduced by the STARFLAG project can be used in many other instances of collective animal behaviour, and it is therefore reasonable to expect that a new generation of 3D data is about to come. In these conditions, it will be possible to make a direct comparison between empirical data and numerical models. Such a comparison will be based on the measurement of well-defined quantities in the two different data sets, empirical and theoretical. Such quantities must be at the same time informative about the properties of the group, and suitable to empirical data. The aim of this work is to present in full detail the quantitative tools of analysis that have been applied to the STARFLAG data set (Ballerini et al. 2008a, b). We stress that these tools can be used on any 3D data set, including those coming from numerical models. We hope this is a step towards standardizing the 3D methods of analysis in collective animal behaviour, making thus smoother the comparison between theory and experiments.

The paper is divided into two parts. In the first part we will deal with the problem of the bias introduced by the individuals at the border of the group. There are well-known statistical methods to solve this problem. Such methods, however, may not be familiar to a biological audience, and thus we will review these methods in a very pedagogical way. In the second part of the paper we will present the STARFLAG tools for the analysis of 3D animal groups. As physicists, we have a very different background to most researchers working in collective animal behaviour, and we think that a cross-fertilization of different concepts and tools may be useful. In this second part we will also illustrate, by using practical example, how the border bias can completely modify a measure, and even introduce some artefacts that can be mistakenly interpreted as real biological features.

The Problem of the Border

One of the major problems we face when analysing the 3D structure of a group of animals is that of the border. Whatever species is observed, animals on the border of the group will inevitably introduce a bias into any measurements taken, unless they are properly treated. This problem has been mostly disregarded by previous empirical observations and numerical studies alike. It is only with a very large number of points, say beyond $10^6$ in 3D, that the ratio between surface and volume becomes small enough that border effects become irrelevant. Such numbers are far beyond the reach of empirical observations, and difficult to attain even in numerical simulations. Thus, border effects *must* adequately be taken into account in any 3D analysis.

To explain in what consists the problem of the border, let us start with a simple example. Consider a school of fishes circling clockwise around an empty core, a pattern known as *milling* (Fig.1). Now imagine that one wishes to compute the spatial distribution of nearest neighbours. The (many) individuals on the external border lack neighbours to their left, whereas the (few) individual on the internal border (the core) lack neighbours to their right. If individuals on the border are included in the analysis, one obtains a distribution indicating that fishes have, on average, fewer neighbours to their left. This result is not, of course, a general consequence of the local interaction rules among the fish: in fact, the completely opposite result would be obtained for a school milling counter-clockwise! This is a typical case where disregarding border effects results in the conflation of two levels of analysis that should remain separate; specifically, the morphological level (the mill-like shape of the school), and the behavioural level (individual interactions and nearest neighbour distribution). This problem is always present, whatever the shape of the group, and it can bias our analysis heavily if it is not treated properly. Fortunately the problem is well known in statistics and there are standard methods to cope with it (Stoyan & Stoyan 1994).

In order to cure the bias of the border, the first essential step is to define, or recognize, those animals that belong to the border of the group. Moreover, identifying the border is also crucial to measuring the volume, and thus the density, of the group. In general, defining the volume of an irregular and discrete set of points is not a trivial problem. However, we often want to measure the volume and density, since many biological issues are related to these quantities. In this section we will explain how it is possible to define the border in a rigorous way, and how to take care of the bias it introduces.

Before we begin, we should note one very important fact. As we have said, the weight of the border in a group is proportional to the ratio between surface and volume. This ratio scales as $1/N^{1/3}$ with the number $N$ of animals in the group: for small groups this ratio is large, and relatively many animals belong to the border, whereas in very large groups only a minority of animals is on the border. For example, in a flock of about 20 birds in 3D, all birds belong to the border. In a flock of 100 birds, only ~3 birds are at the interior of the flock. Thus, working with small groups is very inconvenient, and rather uninformative, because, once the birds on the border have been eliminated from the analysis, no animals are left with which to build statistics. Small groups are completely dominated by border effects, and it is therefore very difficult to get unbiased results out of them. For this reason, collecting data on large groups was one of the main objectives of the STARFLAG project. Of course, this does not mean that smaller groups cannot be studied: many animals live in small groups, and it is certainly useful to study them. However, in these cases one should be particularly careful about the problem of the border, and realize that not all quantities can be measured in an unbiased way.

Convex Hull, Non-Convex Groups and α-Shapes

The simplest tool to define the border of a discrete arrangement of points is the so-called Convex Hull (CH). As its name suggests, the CH is an algorithm that gives the list of points that belong to the convex envelope of the original group (see Fig. 2), either in two or in three dimensions. Apart from giving the border's points, the algorithm also provides the volume of the hull, so that an estimate of the density can readily be calculated. The great advantage of the CH method is that it does not need any adjustable parameters in order to get results and it is, therefore, fast and unambiguous. Moreover, there are several ready-to-use CH routines available on the Internet.

The downside of the method is, of course, that animal groups are not necessarily convex. For example, starling flocks often display concavities in their shape. A school of milling fishes is another obvious example. In non-convex cases, the CH is unable to detect the concavities and its volume includes substantial regions void of points (see Fig. 2). This is very bad for the analysis and, in particular, it introduces a significant bias into all indicators related to the volume and density of the group, and to nearest neighbour distance. The calculated volume may be considerably larger than the real one, so that density will be badly underestimated, and the average nearest neighbour distance overestimated.

We thus need a method that can detect concavities in an aggregation. The first thing to understand when analysing this problem is that there is no absolute way to define the non-convex border of a set of points. Rather, we need to specify the minimum size, or radius $R$, of the concavities we want to eliminate from the surface of the aggregation. Consider Fig. 2: there are two very obvious concavities, one with radius $R_1$ and one with radius $R_2 < R_1$. In order to define a non-convex border, we have to specify whether we want to 'carve' out of the surface of the smaller concavity alone, or eliminate both of them.

Once we have specified the minimum size of the concavity we want to detect, we can use an algorithm called α-Shape (AS in the following). The basic idea of the AS algorithm (Edelsbrunner & Mucke 1994) is to carve the set of 3D (2D) points with spheres (discs) of radius, $R$, where the scale, $R$, is a parameter specified by the user. Whenever the surface (perimeter) of a sphere (disc) hits three (two) points, these points are included in the border. No points are permitted to be within the carving sphere (disc).

In this way, all concavities larger than $R$ are detected (see Fig. 2), and the effect is to obtain the exact set of points belonging to the non-convex border on scale $R$. Note that the convex hull can be obtained as a limiting case of the AS, simply by setting the radius $R$ to infinity (Fig. 2).

Fixing the Scale of the Concavities

The good thing about the AS algorithm is that it is very intuitive to use. The only problem is how to choose the scale, $R$, of the concavities. Clearly, too large a value will neglect some concavities, whereas too small a value will make the spheres penetrate into the aggregation, with the rather odd result that all points will belong to the border. How do we choose $R$? One possibility is to use a biological criterion: for example, $R$ should be significantly larger than the nearest neighbour distance, otherwise the group fragments into many sub-parts. However, these kinds of criteria are normally not sufficient, and they can only provide some guidelines. A less ambiguous way must thus be found.

A good way to fix $R$ is to check the resulting density of the aggregation as a function of $R$. Let $B$ be the number of points belonging to the border and $I$ the number of points in the interior of the aggregation, such that $B+I=N$, the total number of points. Let $V$ be the volume contained within the border defined by the AS algorithm. When $R$ is decreased, more points are included in the border, so that $B$ grows while $I$ decreases. At the same time the volume, $V$, decreases, because more concavities are carved out of the surface. If we plot the density of internal points, $I/V$, as a function of $R$, we discover something interesting: whenever $R$ becomes smaller than the typical scale of a concavity, the internal density makes a sudden jump, because whenever we cut a *bona fide* superficial void, the volume decreases much more steeply than the number of internal points (Fig. 3). From this plot, we can obtain a very clear idea of the sizes $R_1 < R_2 < R_3 < \ldots$ of the concavities in the aggregation, and we can accordingly fix $R=R_1$, i.e. to the smallest genuine concavity. This was the STARFLAG method to fix $R$.

Even though measuring the internal density as a function of $R$ for all 3D groups in the data set may be time consuming, there are no better alternatives. One point worth noting, however, is that, most of the time, the range of possible values of $R$ can be reduced drastically by using biological constraints and experience, so that trying a handful of values of $R$ is normally good enough to eliminate the most obvious concavities. Moreover, in many instances, the shape of an animal group changes rather smoothly across time. This means that, in a given time series, $R$ does not change sharply, so that it can be updated after a long interval. However, whenever the density of the group shows some unexpected or sudden variation, it is always wise to check whether or not this is due to a concavity, whose size is below the chosen value of $R$.

How to Cure the Border's Bias

Once we have defined animals belonging to the border, we must treat them properly, in order to avoid border effects in our final measurement. In statistics, this amounts to computing 'edge-corrected' estimators for the observables one is interested in (for a deeper perspective on this problem see Stoyan & Stoyan 1994).

The basic origin of the border's bias is that border points have on average a very different neighbours statistics, compared to inner points. This is simply due to the fact that a large proportion of the space around them is empty. This is why, for example, the typical distance of the nearest neighbour of a point on the border is significantly larger than the average nearest neighbour distance of points in the interior. Moreover, as the milling fish example illustrates, border points carry a strong signature of the group's shape, which will bias all orientation quantities. Neighbour-based analyses enter to many different observables, which is why border effects are so widespread and so dangerous.

The most elementary method for dealing with border points is the following: consider them as neighbours of inner points, but discard them as focal (or reference) points (Fig. 4). For example, assume that point $P_1$ belongs to the border and that it is at distance 0.8m from point $P_2$, which is an inner point. Assume also that $P_2$ is the first nearest neighbour of $P_1$, and that $P_1$ is the second nearest neighbour of $P_2$ (nearest neighbour relationships are not necessarily symmetric). Then, the value 0.8m is included in the statistics of second nearest neighbours (when considering $P_2$ as focal point), but it is not included in the statistics of first nearest neighbours. This method is quite rough, because it very crudely separates those points that belong to the border from those that do not. In reality, points that are very close to the border without, however, belonging to it, may show biased statistics as well. This is not taken into account by the method just described.

A more precise technique was proposed by Hanisch (1984) in the context of the nearest neighbour distance statistics (see also Stoyan & Stoyan 1994). In order to use this method we must be able to compute, for each given point $i$, the distance between this point and the border, let us call it $D_i$. This is defined as the distance between $i$ and the closest geometric face of the border (Fig. 4). Consider a neighbour $k$ of point $i$ and compute their mutual distance $d_{ik}$. The Hanisch method states that if $d_{ik} > D_i$, then point $k$ must not be included in the statistics of the neighbours of $i$ (Fig. 4). The basic idea is that if $k$ is more distant from $i$ than the border, then their mutual neighbouring relationship is biased by the presence of the border. This method is very simple to employ, it gives very good results, and can be safely used for most observables. When it comes to the distribution of nearest neighbour distances, however, this method slightly overweights very close neighbours (see later). This small problem can be cured by appropriately reweighting all distances. This is the weighted Hanisch method (Hanisch 1984; Stoyan & Stoyan 1994), which was the method used by STARFLAG. Both the simple and the weighted Hanisch estimators are proven to be asymptotically (i.e. for large number of points) unbiased estimators for the nearest neighbour distance distribution. The results obtained with the different methods are given below.

Although correcting for border effects is quite technical and time consuming, it is very important to understand that disregarding these effects often produces disastrous and misleading results. It is impossible to overemphasize this point. In order to illustrate the dangers related to this problem, and how it can produce some very nasty artefacts, we provide some examples of border bias later in the paper.

What to Measure and How to Do It

In this section we list some useful observables, and illustrate some of the subtle technical pitfalls one may encounter. It is clear, however, that the decision of what to measure depends on the scientific aims of a study. In no way do we wish to suggest here that there is a unique, nor a best, way to perform the analysis of 3D data.

<u>Angular Distributions and the problem of the Jacobian</u>

One of the most interesting things to observe in an animal group is the average angular distribution of nearest neighbours with respect to the direction of motion. Neighbour orientation has been used in the past to characterize 3D fish schools (Cullen et al. 1965; Partridge et al. 1980), and, more recently, it has been used in 3D bird flocks as an indirect way to study the range of the interaction among the individuals (Ballerini et al. 2008a). In both fish and birds, the clearest indicator of a non-trivial spatial structure of individuals was directly encoded in the angular distribution of neighbours.
      The measurement proceeds as follows. We identify the nearest neighbour of a focal individual, and, irrespective of the distance between the two, we define the nearest neighbour's vector, **u**. We also define the vector of the velocity **V**. In the following, both **u** and **V** will be normalized, unitary vectors. Note that the velocity, **V**, can be either the actual velocity of the focal individual, or the global (centre of mass) velocity: fluctuations in the orientation of the individuals are normally so small that this does not make any practical difference for what follows. Then we measure the orientation of the nearest neighbour's vector **u** with respect to the velocity **V**. There are two ways of doing this, and we explain both of them below. The important point, however, is that the practical outcome of measuring the orientation of **u** with respect to the **V** is an angle (or a quantity directly related to it). Such angle is then measured for all focal points in the group, and the probability distribution (i.e. the normalized frequency) of this angle is calculated. Finally, this probability distribution must be compared to some null case, and the simplest one is the case where there is no structure at all, and the arrangement of points is completely isotropic in space. This is the case of a spatially random set of points (spatial Poisson process). A significant difference between the measured probability and the Poissonian case indicates that there is non-random, anisotropic structure in the group.
      As we said, there are two different ways to measure the orientation of the nearest neighbour's vector **u**, with respect to the velocity, **V**. First, we can consider the angle, θ, between these two directions, typically by computing the inner (or scalar) product between these two unitary vectors,

$$\cos(\theta) = V_x u_x + V_y u_y + V_z u_z \quad . \quad (1)$$

The angle θ is sometimes mistakenly called 'bearing' angle, but this is inaccurate. We shall see below that the correct bearing angle is, in fact, defined differently. The crucial, and sometimes misleading, point about θ is that, even in a completely random arrangement of 3D points, its probability distribution is not constant. This is due to the presence of a Jacobian factor (in fact, a determinant) in the transformation between

Cartesian and spherical coordinates (Apostol 1969). To understand the origin of the Jacobian factor in the probability distribution of θ, let us consider a sphere of radius one, with the unitary vector **V** corresponding to the north pole. Now consider the circle defined by all points on the sphere that lie at an angle θ with **V**. The size of such circle is proportional to sin(θ): this is the Jacobian factor. The Jacobian is simply telling us that, even in the random case, where the sphere is uniformly covered by points, there are many more points on the equator (θ=π/2, sin(θ)=1) compared to the poles (θ=0, θ=π, sin(θ)=0). The probability distribution of θ in the random case is thus given by:

$$p_{random}(\theta) = \frac{1}{2}\sin(\theta) \quad ; \quad \theta \in [0:\pi] \quad , \quad (2)$$

where sin(θ) is the Jacobian. It is therefore highly inconvenient to compare the measured distribution of θ with the random one, since this is not a constant. The comparison between two non-constant and certainly noisy functions can be difficult.

The best thing to do is to deal directly with the cosine of θ, cos(θ)=y. Indeed, the distribution of y, unlike that of θ, is flat, since there are two Jacobian factors that cancel each other out:

$$p_{random}(y) = p_{random}(\theta)\left|\frac{d\theta}{dy}\right| = \frac{1}{2}\sin(\theta)\frac{1}{\sin(\theta)} = \frac{1}{2} \quad , \quad (3)$$

with:

$$y = \cos(\theta) \in [-1:1] \quad , \quad (4)$$

and it can be conveniently compared to the empirical data. In fact, in many biological cases $p(y)$ is non-constant, signifying a non-isotropic arrangement of neighbours.

A different, and more complete, way to characterize the orientation of the nearest neighbour vector **u** in space is to fix a 3D reference frame, and to associate two angles to the nearest neighbour vector **u**, rather than simply one. To explain how this is done, we take birds as a reference system. First, we consider the plane containing the velocity **V** and the wings of the focal bird, i.e. the coronal plane (Fig. 5). Second, we consider the direction orthogonal to the coronal plane, i.e. the dorsal-ventral direction: we call **G** the unit vector along such direction. (In most situations, and especially in fish and non-turning birds, this direction is almost parallel to gravity, but this is not essential for its definition). Now we can define two angles (Fig. 5): the bearing angle, α, is the angle between the velocity, **V**, and the projection of u onto the coronal plane; the elevation, ϕ, is the angle between u and its projection onto the coronal plane. It is important to understand that the bearing angle, α, is different from the angle θ described above. In fact, unlike θ, the probability distribution of the bearing angle is constant,

$$p_{random}(\alpha) = \frac{1}{2\pi} \quad ; \quad \alpha \in [-\pi:\pi] \quad . \quad (5)$$

On the other hand, the distribution of the elevation angle ϕ contains a Jacobian, and it is therefore not a constant:

$$p_{random}(\phi) = \frac{1}{2}\cos(\phi) \quad ; \quad \phi \in [-\pi/2:\pi/2] \quad . \quad (6)$$

The bottom line of this section is that, whenever we want to investigate the angular distribution of neighbours (nearest neighbours, but also second nearest, third nearest and so on), by plotting the probability distribution of some quantity, we must make sure to choose a quantity whose distribution is constant in the trivial random case. The best choice is either cos(θ), where θ is the angle between velocity and neighbour, or the bearing angle α, which is the angle between the velocity and the projection of the neighbour's direction onto the coronal plane. Both these quantities have a constant probability distribution, so that any non-constant trend of their distribution is a clear indication of non-isotropic structure of neighbours.

Quantifying the Anisotropy

All empirical studies that have investigated the orientation of neighbours in animal groups have detected a strong anisotropic structure (Cullen 1965; Partridge et al. 1980; Ballerini et al. 2008a, b). This means that the direction of motion breaks the rotational symmetry of the group's structure. In other words, by taking the velocity as the reference direction, animals are located with respect to one another in some preferential direction. The angular distributions discussed above show this fact very clearly. However, it is sometimes useful to quantify the strength of this anisotropy by means of a single scalar quantity, rather than a graph, or a function. To this end, STARFLAG introduced a novel tool, the anisotropy factor γ (Ballerini et al. 2008a).

Given a certain group, consider the set of all nearest neighbour vectors, $\{\mathbf{u}^i\}_{i=1...N}$ where $N$ is the number of individuals, and each vector is normalized to unity. For each one of these vectors, $\mathbf{u}^i$, we can build the following projection matrix,

$$\mathbf{M}^i = \begin{pmatrix} u_x^i u_x^i & u_x^i u_y^i & u_x^i u_z^i \\ u_y^i u_x^i & u_y^i u_y^i & u_y^i u_z^i \\ u_z^i u_x^i & u_z^i u_y^i & u_z^i u_z^i \end{pmatrix} \quad \text{or, equivalently,} \quad \mathbf{M}_{\alpha\beta}^i = u_\alpha^i u_\beta^i \quad , \quad (7)$$

where α,β=x,y,z. The effect of applying this matrix to an arbitrary vector $\mathbf{v}$ is to produce a new vector, which has the same direction as $\mathbf{u}^i$ and modulus equal to the projection of $\mathbf{v}$ along the direction of $\mathbf{u}^i$. Thus, the matrix projects along the direction of $\mathbf{u}^i$. Indeed,

$$\sum_\beta \mathbf{M}_{\alpha\beta}^i v_\beta = u_\alpha^i \sum_\beta u_\beta^i v_\beta \quad ; \quad \alpha,\beta = x,y,z \quad . \quad (8)$$

For this reason $\mathbf{M}^i$ is called projection matrix. We can now average this matrix over all neighbours in the system, to obtain a matrix that projects along the average direction of the nearest neighbour:

$$\mathbf{M}_{\alpha\beta} = \frac{1}{N} \sum_{i=1}^{N} u_\alpha^i u_\beta^i \quad . \quad (9)$$

The important point here is that the three eigenvectors of the average projection matrix, $\mathbf{M}$, correspond to the most relevant directions in space. In particular, the eigenvector corresponding to the smallest eigenvalue of $\mathbf{M}$ indicates the direction along which there is the smallest number of nearest neighbours vectors, and it therefore indicates the direction along which, on average, one is less likely to find an animal's nearest neighbour. Let us call this (unitary) eigenvector, $\mathbf{w}$. In a random, isotropic arrangement of points, $\mathbf{w}$ lies in a random direction: there will always be a direction along which, by random fluctuations, the density of nearest neighbours will be slightly smaller than the average. Therefore, in an isotropic group, we do not expect $\mathbf{w}$ to have any particular correlation with the velocity $\mathbf{V}$. Conversely, whenever there is some anisotropic, non-random structure of neighbours, the minimal crowding direction $\mathbf{w}$ will have a preferential direction with respect to the global velocity **V**. This can be quantified by measuring the square inner product between these two vectors,

$$\gamma = \left(\sum_\alpha V_\alpha w_\alpha\right)^2 \quad ; \qquad \gamma \in [0:1] \quad . \tag{10}$$

This is the anisotropy factor: a value of γ close to 1 indicates that the direction of minimal crowding of the nearest neighbours is almost parallel to the velocity. A value of γ close to 0 indicates that the direction of minimal crowding is orthogonal to the velocity (this does *not* imply that the velocity is the direction of maximal crowding). To fix a scale, let us compute the value of γ for the case where there is no correlation whatsoever between the two directions. To average γ in the random case we can use spherical coordinates, with the velocity **V** along the *z*-axis,

$$\gamma_{random} = \frac{1}{4\pi} \int_{sphere} dx\, dy\, dz\, (\mathbf{V} \cdot \mathbf{w})^2 = \frac{1}{4\pi} \int_0^{2\pi} d\varphi \int_0^\pi d\theta\, \sin(\theta) \cos^2(\theta) = \frac{1}{3} \tag{11}$$

Thus, if we measure γ many times for two uncorrelated directions, we get, on average, the value 1/3. This value can be taken as a benchmark for a complete decorrelation. Of course, even in the random case, in a single instance, the value of γ can take any positive value; however, if we average it over many instances, we will get 1/3, signifying that there is no anisotropy in the system. On the contrary, if γ is significantly larger than the isotropic, random value of 1/3, then there is correlation between global velocity and direction of minimal neighbour crowding. Within STARFLAG, we discovered that γ is systematically larger than 1/3 for nearest neighbours, indicating that there is a significantly lower probability to find nearest neighbours along the direction of motion (Ballerini et al. 2008a).

      The useful thing about the anisotropy factor, γ, is that it can be calculated as easily for nearest neighbours, as for second nearest neighbours, third nearest neighbours, and so on. Of course, in doing this, one should always be very careful to take care of border effects, and use the Hanisch method to eliminate this bias. We calculated the anisotropy factor as a function of the distance of the neighbour from the focal individual, and we found that γ decreases with this distance. This allowed us to define a range of the anisotropy, as the point where γ becomes close to 1/3. This range can be computed either in units of metres (metric distance), or in units of individuals (topological distance). What we found that, for starlings, the most appropriate distance measure is topological, not metric: the topological range is the same in flocks with different densities, whereas the metric range depends on density, scaling linearly with the average nearest neighbour distance (Ballerini et al. 2008a).

      Note that the anisotropy factor we have defined refers to the direction of minimal crowding. Of course, one can also consider the direction of maximal crowding, by computing the square inner product between the velocity and the eigenvector relative to the largest eigenvalue. In the case of starlings, the direction of maximal crowding, unlike the minimal crowding direction, does not have any clear correlation with velocity. For this reason, we defined the anisotropy factor using the direction of minimal crowding. However, it may be that, in different systems, the direction of maximal crowding gives a sharper description of the group's anisotropy. In general, one should

always compute the inner products of the velocity with all three eigenvectors of the anisotropy matrix.

Nearest Neighbour Distance and Exclusion Zone

The second most investigated observable, which is not related to orientation properties, is nearest neighbour distance (NND); in particular its probability distribution. Apart from its obvious biological significance, this function is very useful for providing a quantitative estimate of the so-called "exclusion zone" around each individual.

Almost all models of collective animal behaviour assume that, surrounding each individual, there is a zone where other individuals are not found. This can be modelled in different ways, either with a strong short-range repulsion, giving rise to what is known as a soft-core, or alternatively it can be modelled with a *bona fide* hard-core, i.e. an impenetrable region surrounding each individual. (In this last case, if we call $r_0$ the hard-core, the minimum distance between points is, of course, $2r_0$.) The existence of an exclusion zone is a reasonable biological assumption, and one would like to measure the radius of this zone in real biological systems and then feed this value into model simulations. This can be done using the distribution of nearest neighbour distances.

The first thing to note here is that, like everything else, real exclusion zone sizes fluctuate across individuals. This means that, in order to obtain the typical value of the exclusion zone, $r_0$, we cannot simply measure the minimum nearest neighbour distance, $d_{min}$, within a large group and claim that this is twice the exclusion zone: the value of $d_{min}$ invariably refers to a pair of animals that are extremely close to each other, due to some rare fluctuation. In other words, in real biological groups, the exclusion zone is not strictly hard, so we cannot simply measure the minimum distance.

A simple way to find the typical size of the exclusion zone is to fit the nearest neighbour distance distribution using the exact equations describing the same quantity for an arrangement of totally impenetrable hard spheres ("hard-spheres"). Hard spheres are a simple and popular test-bed system in physics: the only interaction they have with each other is due to their hard exclusion zone of radius $r_0$, i.e. the hard-core. Thus, they behave very much like a set of billiard balls hitting each other. Of course this is a quite crude approximation of animal groups. However, the hard-spheres system has the great virtue of simplicity, with two important by-products: first, a hard-sphere system is described by only two parameters, i.e. density $\rho$ and hard-core $r_0$; second, there are numerous exact results for hard-spheres, that can be directly compared to measured quantities. Therefore, as long as we use hard-spheres only to describe the short distance properties of the group, and in particular to fit the exclusion zone, we should not be concerned about oversimplifying our analysis.

Rather than dealing with the nearest neighbour distance distribution, $P(r)$, it is more convenient to use the cumulative distribution, $P^>(r)$, defined as the probability that two nearest neighbours have mutual distance larger than $r$. The former is simply the derivative of the latter. Note that, by definition, $P^>(r)=1$ for $r < 2r_0$ in hard-spheres. For $r > 2r_0$ the exact expression of $P^>(r)$ in a hard-spheres systems with density $\rho$ and hard-core radius $r_0$, is given by the following (equation 5.96 of Torquato 2002):

$$P^>(r) = \exp\left\{-A\left[\left(\frac{r}{2r_0}\right)^3 - 1\right] + B\left[\left(\frac{r}{2r_0}\right)^2 - 1\right] - C\left[\left(\frac{r}{2r_0}\right) - 1\right]\right\} \quad , \quad (12)$$

where,

$$A = 8\,\phi(1 + 4\phi) \quad ,$$
$$B = 18\,\phi^2 \quad ,$$
$$C = 24\,\phi^3 \quad ,$$

and,

$$\phi = \frac{4}{3}\pi r_0^3 \rho \quad , \quad (13)$$

is the so-called packing fraction, i.e. the ratio between the total volume occupied by the spheres and the volume of the group. The expressions of the constants $A$, $B$, and $C$ are given at the first order in $\phi$, and thus are valid only in the limit of small packing fractions, which is often verified in many animal groups (for example, in starling flocks; Ballerini et al. 2008b). Equation (12) can then be used to fit the cumulative nearest neighbour distribution (in fact, it is technically more convenient to fit the logarithm of $P^>(r)$). The parameters of the fit are the density $\rho$ and the radius of the exclusion zone $r_0$. Note that a random set of points with given density (spatial Poisson point process) can be obtained in the obvious limit $r_0=0$, so that formula (12) can also be used to fit the data to the Poisson case, where the cumulative distribution takes the simpler form $P^>(r)=\exp\{-(4/3)\pi\rho r^3\}$. Finally, we note that the hard-core fit can also be used to give an estimate of the compactness of the aggregation. In order to do so, one has to look at the packing fraction, $\phi$, of the aggregation: small values of $\phi$ correspond to sparse systems, while large values indicate compact ones. As reference values, for hard spheres the crystalline arrangement corresponds to $\phi = 0.49$, and the random close packing to $\phi \sim 0.79$ (Torquato 2002).

In Ballerini et al. (2008b) we present data for the nearest neighbour distribution for starling flocks, and show that a hard-spheres fit is significantly better than the Poisson case fit. In particular, it is evident that the empirical distribution has a lack of nearest neighbours at very short distances, compared to the Poisson. Clearly, however, the hard-sphere fit is, generally speaking, far from perfect, because, as already mentioned, the exclusion zone is not strictly hard. However, it does a much better job than the Poissonian case, so that, as a first approximation it can be used as a reliable estimate of $r_0$. More refined models can be used, of course, with the advantage of increased realism, but the disadvantage of a larger number of parameters: as always, one must carefully balance the trade-off between simplicity and realism.

What Happens When the Border is Disregarded: Instructive Examples

In this section, we will present three practical examples of how disregarding the border can produce completely misleading results. In all cases. we consider aggregations of points that are randomly generated in 3D with a given density: what is known as a spatial Poisson point process (for a more complete description of Poisson point fields see Stoyan & Stoyan 1994). The Poisson point process is the prototype for a spatially homogeneous and isotropic arrangement of points. All observables are in this case known analytically, so that any measurement taken must be consistent with the exact formulas. For this reason, Poissonian aggregations represent an ideal null model with which to test our analysis tools.

In previous studies, both empirical and simulations, the average NND of animal groups is reported as decreasing with an increasing number of individuals within the group (see, for example, Partridge et al. 1980). This implies that animals are more closely packed in a large aggregation than in small ones. STARFLAG did not find any evidence of this behaviour in our large starling flocks (Ballerini et al. 2008b). In fact, we believe that this result is a pure artefact due to a border bias. In Figure 6, we plot the average NND of a spherical aggregation of random points vs. the number of points $N$. All aggregations of different sizes are generated with the same density, and thus they should all display the same NND. However, as we have already noted, points on the border have, on average, more distant nearest neighbours for purely geometric reasons, so that they bias the NND towards larger values if they are included in the statistics. We did *not* compensate for this border effect in Fig. 6, and we computed the NND using all points: the net effect is that NND strongly decreases with $N$, at least for values of $N$ smaller than 100. Only for very large values of $N$ is the surface-to-volume ratio small enough to suppress the bias introduced by the border points, and the NND converges to its correct value. Thus, this packing effect as a function of aggregation size has no biological origin. It is simply an artefact of failing to account for the bias introduced by border points.

As a second example, we consider the cumulative distribution of the nearest neighbour distance, $P^>(r)$. For a Poisson aggregation, the exact value is given by equation (12), with $\phi=r_0=0$, that is $P^>(r) = \exp\{-(4/3)\pi\rho\, r^3\}$. In Fig. 7 we plot the logarithm of $P^>(r)$ as a function of $r$. It is clear that including border points in the calculation gives a very bad estimate of $P^>(r)$, compared to the exact case. The rough method for eliminating border points (described above) significantly improves the estimate. The simple Hanisch method improves the empirical estimate even further, and with the weighted Hanisch method one gets an excellent consistency with the exact result.

As a final example, we can show that the orientational properties of groups can be badly biased by border effects. We considered a Poisson point process of random points generated within a relatively thin 3D slab, with an aspect ratio similar to a typical starling flock (1:3:7). We put this in motion first along its shortest axis, and then along its longest axis (Fig. 8). Given that the points are random, there should be no correlation between the velocity and the eigenvectors of the anisotropy matrix, so that by repeating the numerical experiment many times we should find that the anisotropy factor $\gamma = 1/3$, not only for the first nearest neighbours, but also for the second, third and so on. However, when border effects are disregarded, this is not the case (Fig. 8), simply because, depending on the direction of motion, border points will have more or less neighbours in the direction of the velocity. Only by using the Hanisch method can we recover the exact value of 1/3. As with milling fish schools, a failure to account for border points means that, effectively, we are conflating two very different results: structure (the fact that there is no preferential directions of neighbours) and morphology (the fact that the slab has very uneven dimensions). These two levels, however, must be kept separated.

A final remark is in order. We do not claim here that border effects are exclusively produced by geometry (or physics), and not by biology. In fact, it is reasonable to expect that individuals at the border of the group do behave differently compared to others, for a number of biological reasons. It is exactly to detect in the clearest possible way these biological effects, however, that one must first get rid of the trivial geometric effects. If this is not done, important biological properties can be hidden or distorted by trivial geometric features.

## Main Axes, Dimensions and Aspect Ratios

Even though this work was dedicated to obtaining and analysing the 3D positions of individual animals within a group, there are numerous global quantities that are worthy of investigation. In particular, the morphological properties of a group and its orientation in space with respect to gravity and velocity are of obvious relevance.

The first step is to define the main axes of the group. Our method is simple. We start by fitting the 3D group of points to a 2D plane, let us call it $\Pi$; the direction orthogonal to this plane defines the first axis, $I_1$. We then project all points onto the plane $\Pi$, and we fit such a 2D set of points to a line, let us call it $\lambda$; the line belonging to $\Pi$ and orthogonal to $\lambda$ defines the second axis $I_2$. Finally, $\lambda$ itself defines the third axis $I_3$. This is a set of orthogonal directions in space, and its definition simply requires a linear fit, and it is thus very fast. Of course, if the group is spherical, the two fits are ill defined, so that both the plane and the line will be arbitrarily defined. This result is, in fact, correct, because in the case of a sphere there are no preferential axes, and one can choose an arbitrary set of orthogonal vectors.

Whenever the typical shape of a group is non-spherical, it is important to check whether there is any particular orientation in space. There are two obvious preferential directions in 3D animal groups, namely gravity G and velocity V, and it is thus natural to measure the inner products between the three axes $I_1 I_2 I_3$, and both G and V. For example, in starling flocks the shortest axis, $I_1$, is always more or less parallel to gravity and orthogonal to the velocity, whereas the axes $I_2$ and $I_3$ are approximately orthogonal to gravity, but show no particular correlation with the velocity. Finally, it is interesting to measure the angle between gravity and velocity: intuitively we expect animals to fly, or swim, while remaining as level as possible, in order not to waste potential energy. This is confirmed in starlings by the fact that velocity and gravity are always quite orthogonal to one another. Of course, the situation may be different under external stimuli, especially predation, when the mutual orientation of gravity and velocity may change drastically.

The second important global observable is the dimension of the group along each axis: $I_1$ is the smallest dimension, $I_2$ the intermediate one, and $I_3$ the largest dimension, $I_1 < I_2 < I_3$. There are several ways to do this, the easiest of which is to use an algorithm that finds the Minimal Bounding Box (MBB) of the set of points (Fig. 9a). Note that the naive bounding box, computed using a set of arbitrary Cartesian axes is absolutely unreliable (Fig. 9b). The lengths of the sides of the MBB give a reasonable estimate of the group's dimensions for convex shapes. For consistency, one must check that the main axes defined above are similar to the orientations of the MBB.

Unfortunately, whenever the shape of the group is non-convex, the MBB gives an inaccurate estimate of the dimensions (Fig. 9c). Non-convex animal groups are not at all uncommon, and they are frequent among starling flocks. Thus, within STARFLAG, we used a different method to define these dimensions, which requires knowledge of the α-shape. The idea here is that the smallest dimension of a set of 3D (2D) points is given by the diameter of the largest sphere (disc) entirely contained in the set (Fig. 9d). This principle can be applied provided that we are able to distinguish between what is within the group from what is outside the group, and this can only be achieved once the α-shape algorithm has computed the faces of the polyhedron, i.e. the surface of a group. We then proceed in the following way: we compute the smallest distance between each inner point and the surface: this is the radius of the largest sphere centred on that point and contained entirely in the group. We then select the largest among these smallest distances, and use twice this value as an estimate of $I_1$. We then project all points onto the plane $\Pi$ defined above, and find the minimum dimension $I_2$ of such a 2D group using the same principle (with discs, in place of spheres). We finally project the 2D group onto the direction $I_3$ and identify the dimension $I_3$ as the maximal extension along this line.

The dimensions along the three axes give a measure of the aspect ratio of the group. For example, if the group has an approximately spherical shape the three axes have similar values, $I_1 \sim I_2 \sim I_3$. In starlings, on the other hand, we found that $I_1$ is always considerably shorter than the other two dimensions (Ballerini et al. 2008b). To quantify the aspect ratios we simply measure the two major dimensions in units of the smallest one. Thus, for starlings we say that, on average, the aspect ratio is 1:3:6, meaning that $I_2/I_1 = 3$ and $I_3/I_1 = 6$. A very interesting result in starling flocks is that, even though the dimensions of a group may vary considerably, its aspect ratio is relatively stable, suggesting that there is a mechanism operating within flocks that produces an almost invariant aspect ratio (Ballerini et al. 2008b). Similar results have been found in fish schools (Partridge et al. 1980). It would be very interesting to investigate aspect ratios in other species.

Conclusions

In all the methods discussed in this work we put the emphasis on the mathematical rigour. The reason why we did that is not simply because we are physicists, nor because we think that mathematics, or physics, must be predominant in the study of collective animal behaviour. On the contrary, we believe that in order for the fundamental biological properties of animal groups to emerge in the clearest way, it is necessary to wipe out a measurement from the trivial geometrical effects. As we explained through several practical examples, a lack of mathematical rigour can lead to erroneous conclusions, or, in the best case scenario, make our measurements very difficult.

Let us go back to the nearest-neighbour distance. We have proved that, *even in a completely random arrangement of points*, when border effects are neglected the nearest neighbour distance shows a strong dependence on the group size. The fact that this dependence is present in the random case proves that its origin is exclusively *geometrical*. Such effect is thus biologically spurious, and it must be taken away in order to appreciate the true biological aspects. Note that, by doing this, we are by no means excluding that there may be also a *biological* dependence of nearest-neighbour distance on group's size. However, to investigate the biological contribution it is first essential to get rid of the purely geometrical contribution. An identical logical pattern emerges with the problem of the Jacobian: the angular distributions *do have* a nontrivial shape that has a deep biological origin and meaning. In fact, this is one of the most relevant features of animal groups. However, the Jacobian factor, if disregarded, makes it very difficult to distinguish what is a biological feature (present only in animal groups, and therefore important), from what is a purely geometrical feature (present also in random groups, and therefore irrelevant).

Our approach is therefore not motivated by a pedantic, and pointless, obsession for the mathematical details. We know only too well that empirical data are rough, noisy, scarce, and very difficult to obtain. In fact, empirical data are pure gold in the field of collective animal behaviour. Mathematical rigour is necessary to extract from such data the maximum amount of biological information.

Acknowledgements

We warmly thank Frank Heppner for many useful suggestions and for carefully reading the manuscript. The STARFLAG handbook was conceived during the participation of one of us (A.C) to the 2007 International Ethological Conference in Halifax, Nova Scotia. A.C. wants to thank the participants to this conference for many stimulating discussions and for the exciting scientific environment. This work was financed by a grant from the European Commission under the FP6-STARFLAG project.

FIGURES

Fig.1

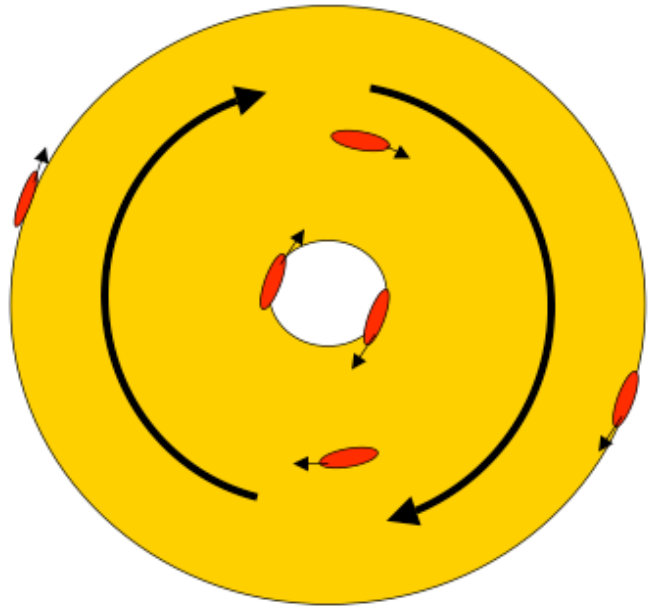

Figure 1. Border bias in a milling configuration
The yellow area represents a school of fish in a milling, or toroidal, configuration: all animals rotate clockwise around an empty core. Animals on the external border lack neighbours on their left, whereas animals on the internal border lack neighbours on their right. However, animals on the external border are much more numerous than animals on the internal border. As a consequence, if we include all animals in the statistics of nearest neighbours, we find that on average animals have a lower probability to have their nearest neighbour to their left. This result would be the opposite for a school rotating counter-clockwise, showing that it is not a fundamental feature of the inter-individual interaction, but rather an artefact due to the undue mixture of structural (nearest neighbour location) and morphological (toroidal shape) results.

Fig.2

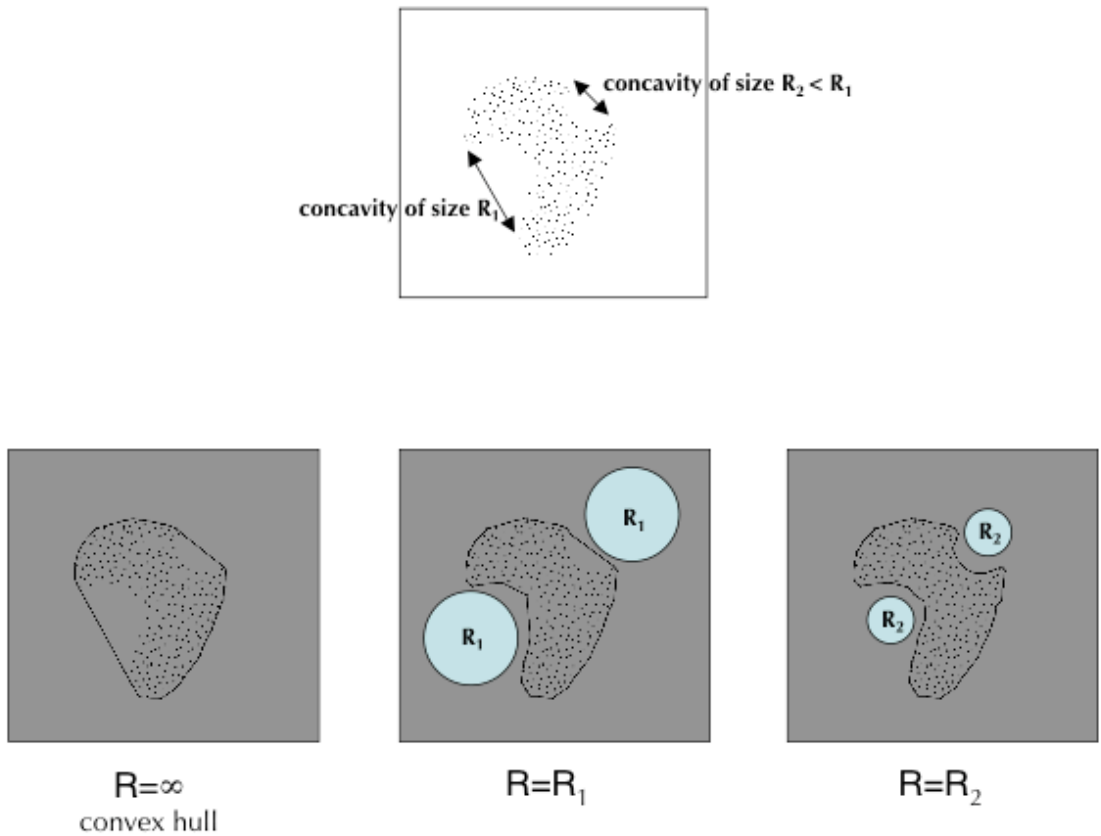

Figure 2. Convex hull and alpha shape
This 2D aggregation of points displays two clear concavities of different sizes. The convex hull of the aggregation includes all points of the convex envelope, and it is by construction unable to detect the concavities. Thus, the convex hull badly overestimates the volume occupied by the aggregation, introducing thus a strong bias in the density of points. Moreover, such a convex border would be unable to properly cure the border effects, since many points effectively on the border are in fact considered as internal points. The alpha shape carves the aggregation with discs of a certain radius (fixed by the user): when a disc touches two points, they are included in the border. In this way the alpha shape detects concavities that have roughly the same size as the chosen radius of the discs.

Fig.3

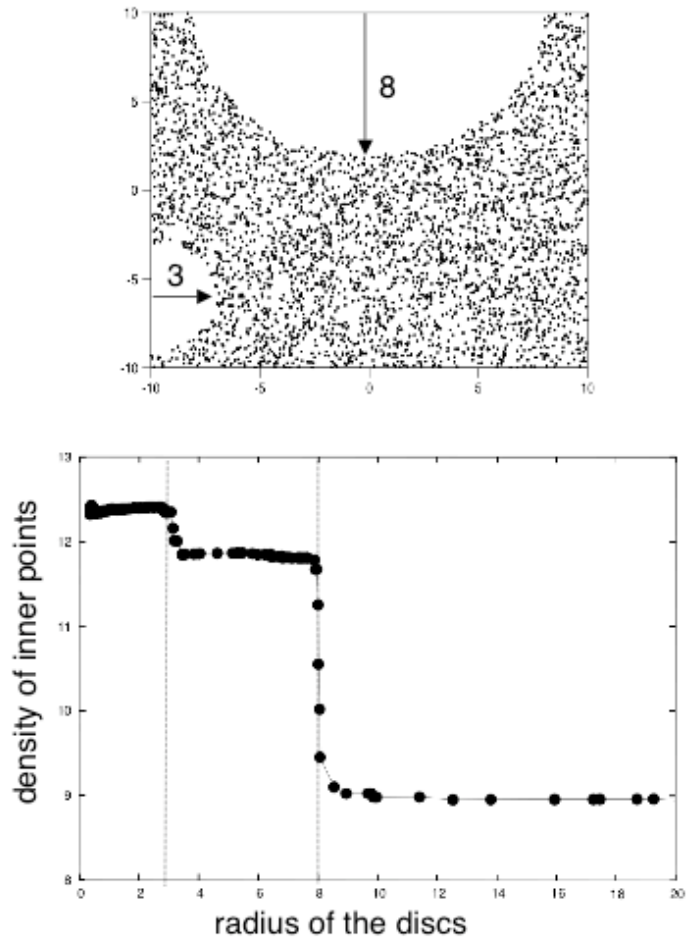

Figure 3. Fixing the scale of the concavities
A 2D aggregation of random points with two clear concavities with different sizes $R_1 = 3$ and $R_2 = 8$. When we plot the density of internal points as a function of the radius $R$ used by the alpha shape algorithm, we find two sudden jumps in correspondence of the radius of the two concavities. In this way it is possible to fix the proper value of $R$ right below the value of the smallest concavity.

Fig.4

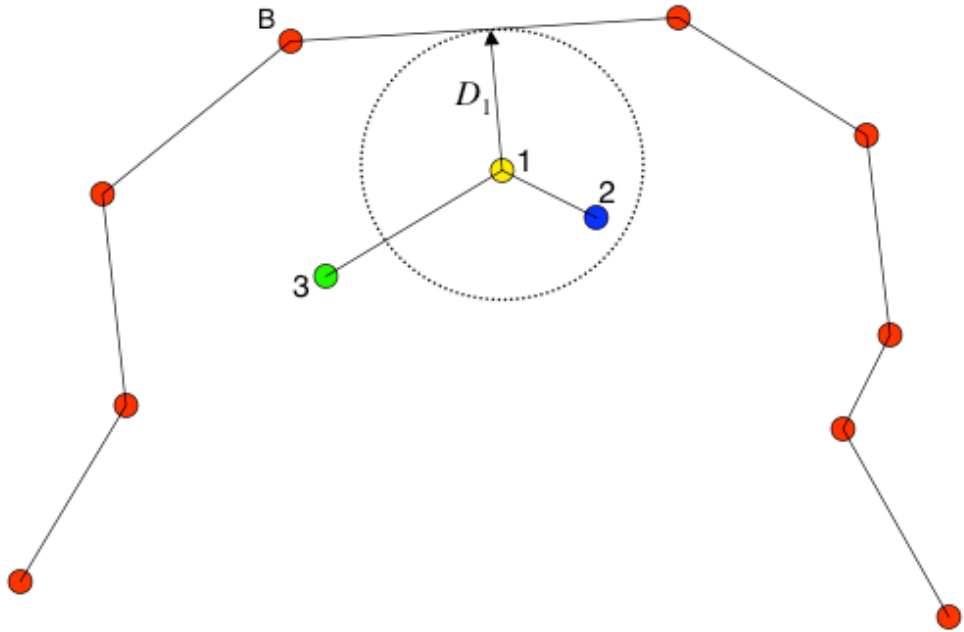

Figure 4. Methods for curing the border's bias

Points on the border have a modified statistics of their neighbours. In particular, the nearest neighbour of a border point is, on average, at a larger distance compared to points at the interior. This is simply due to the fact that half the space around a border point is empty, so that it has fewer neighbours in general. The most basic method to cure the border's bias consists in not using as focal points all points belonging to the border (red in the figure). This means, for example, that when we compute the distances of the neighbours of point 1 (focal point) we can include point B; however, we must not include the distances of the neighbours of point B (focal point) in the global statistics. A more reliable technique is the Hanisch method: given a focal point (1 in the figure), we compute its distance to the border, $D_1$, and only include in the statistics those neighbours of 1 that are closer than $D_1$. Therefore in the neighbours' statistics of focal point 1 we include 2, but not 3. The idea is that on the scale of the distance 1-2, point 1 is for all practical purpose an inner point, whereas on the scale of the distance 1-3, point 1 is influenced by the presence of the border.

Fig.5

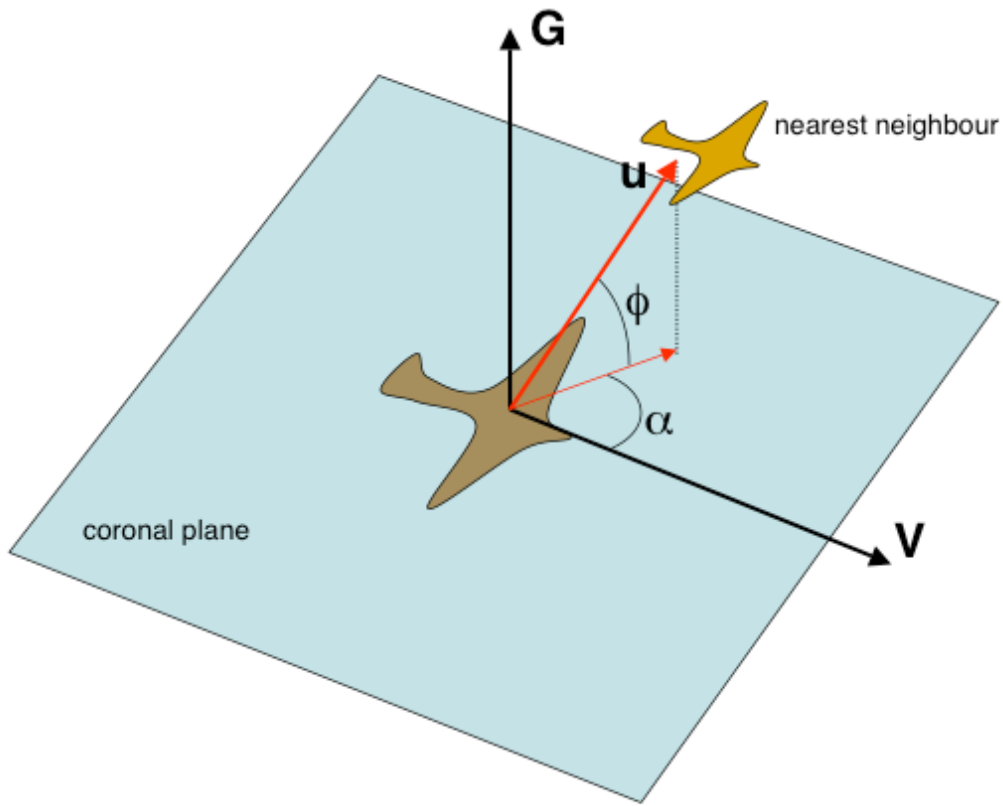

Figure 5. Definition of bearing and elevation
The coronal plane includes the velocity V and, in the case of birds, the wings. The orthogonal direction G to this plane normally is very close to gravity, even though this is inessential to the definition of the reference frame. The nearest neighbour vector u is projected onto the coronal plane: the angle $\alpha$ between this projection and the velocity defines the bearing angle. The angle $\Phi$ between u and its projection onto the coronal plane defines the elevation angle. A third angle, not to be confused with bearing nor elevation, is the angle $\theta$ between u and V (not indicated in the figure).

Fig.6

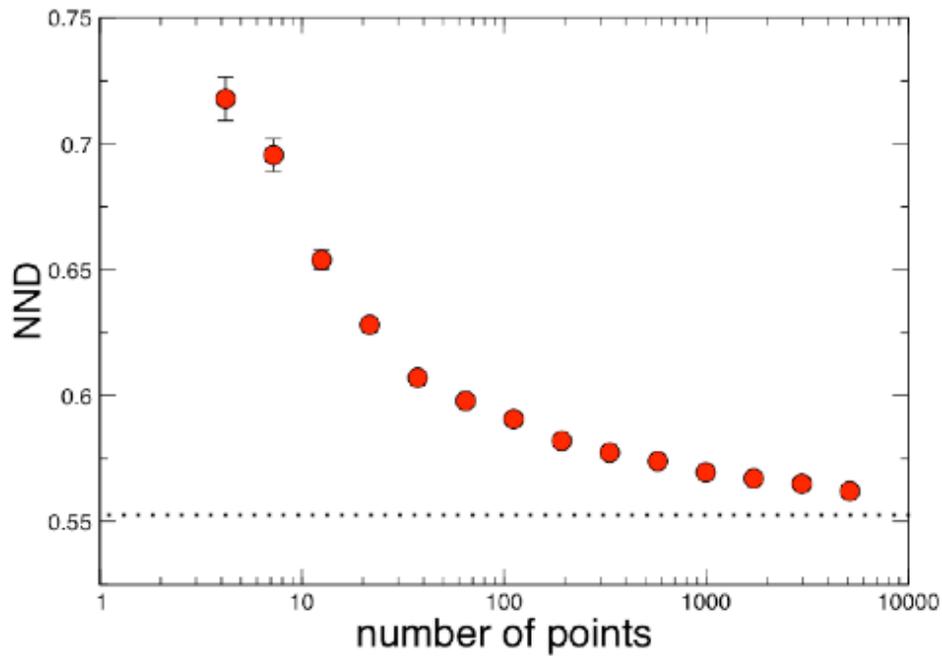

Figure 6. Border artefact I: NND vs. number of points
We generated random aggregations with increasing number of points *N* using a Poisson point process. All aggregations have a 3D spherical shape, and they are all generated with the same density ρ=1, corresponding, by construction, to an average nearest neighbour distance (NND) equal to 0.554 (for the exact relation between density and average NND in a Poisson point process see Stoyan & Stoyan 1994). In other words, the ratio between the number of points and the volume of the sphere within which they are randomly drawn is the same for all sizes. We measure the NND by averaging over all the points of the aggregation, as normally done in the literature for biological groups (from 100 to 10000 samples for each value of N; error bars represent standard error). For small groups size the resulting NND is strongly biased to larger values. The net effect is that, due to the border's bias, the NND decreases as a function of *N*, and only for very large aggregations it reaches the asymptotic correct value. In particular, for *N* < 100 (the typical range of previous empirical studies) there is no hint of an asymptotic value. The decrease of the NND with *N* has nothing to do with any biological feature. It is simply an artefact of having included the border's points in the statistics.

Fig.7

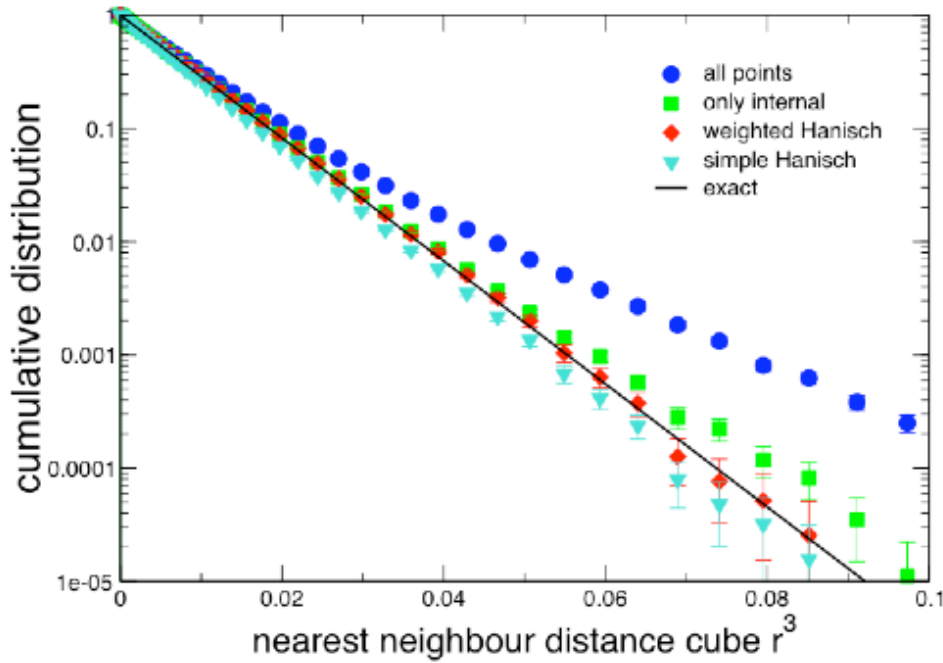

Figure 7. Border artefact II: distribution of nearest neighbours
We generated spherical 3D aggregations of 1000 random points using a Poisson point process, and we compute the cumulative probability of the nearest neighbour distance, $P^>(r)$, i.e. the probability that the NND is larger than $r$. In the Poissonian case the logarithm of $P^>(r)$ is a power law with exponent 3 (see text), so that if we plot $P^>(r)$ vs. $r^3$ in semilog scale the exact result corresponds to a straight line. The black full line is the exact analytic case, which must be recovered by the empirical measure. The blue circles are what we get when we include all points in the statistics, disregarding border effects. The deviation from the exact case is substantial. The green squares are obtained by using the most basic method of including only internal points (see text). This improves a lot the result, even though a significant deviation still persists, especially in the tails. The light blue triangles are obtained by using the simple Hanisch method. The estimate is better than in the previous case, but one can note a tendency to underestimate the distances. The red diamonds are obtained by using the weighted Hanisch method. The result in this case is in very good agreement with the exact curve. Each curve is obtained by averaging over 100 independent samples; error bars represent standard errors.

Fig.8

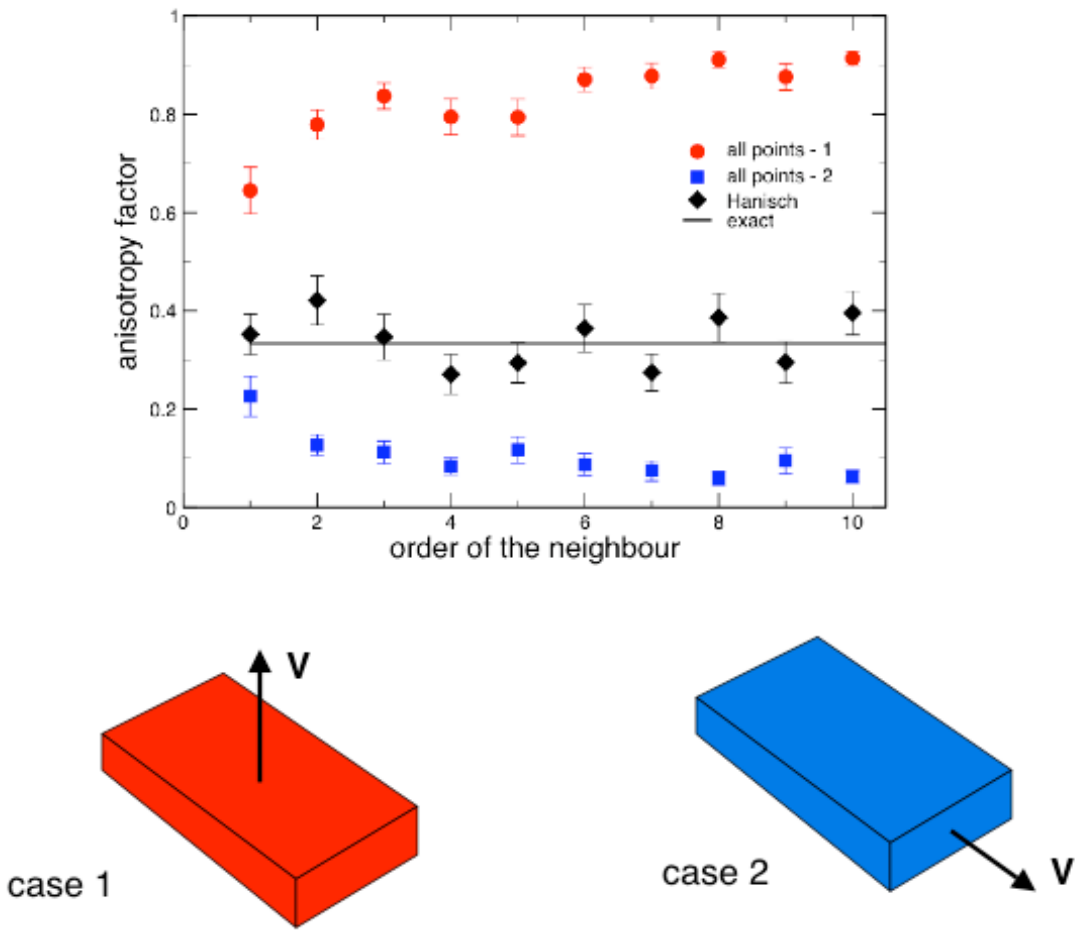

Figure 8. Border artefact III: anisotropy of a moving slab

Anisotropy factor as a function of the order of the neighbour (nearest, second nearest, third nearest, etc.). We generate 1200 random points within a 3D slab of aspect ratio 1:3:7 (similar to realistic flocks of stalings). The spatial distribution of points is random so that there is no intrinsic anisotropy in the distribution of nearest neighbours. We then put in motion the slab along one of its three axes with a velocity V and compute the anisotropy factor $\gamma$ as a function of the order of neighbours. If border effects are disregarded, the value of $\gamma$ strongly depends on the direction of motion. When the slab moves along on of the shortest axis (case 1), the majority of border points lack neighbours along the direction of motion; as a consequence $\gamma$ is larger than the random case 1/3 (red circles). On the contrary, when the slab moves along the longest axis (case 2), the majority of border points lack neighbours in the directions perpendicular to V; as a consequence $\gamma$ is smaller than the random case 1/3 (blue squares). None of these two results is correct, since they both mix some morphological information (the slab proportions) to structural information (the anisotropy factor). When the border's bias is properly cured by using the Hanisch method, the exact value 1/3 is recovered (black diamonds). Points are averaged over 50 samples; error bars represent standard error.

Fig.9

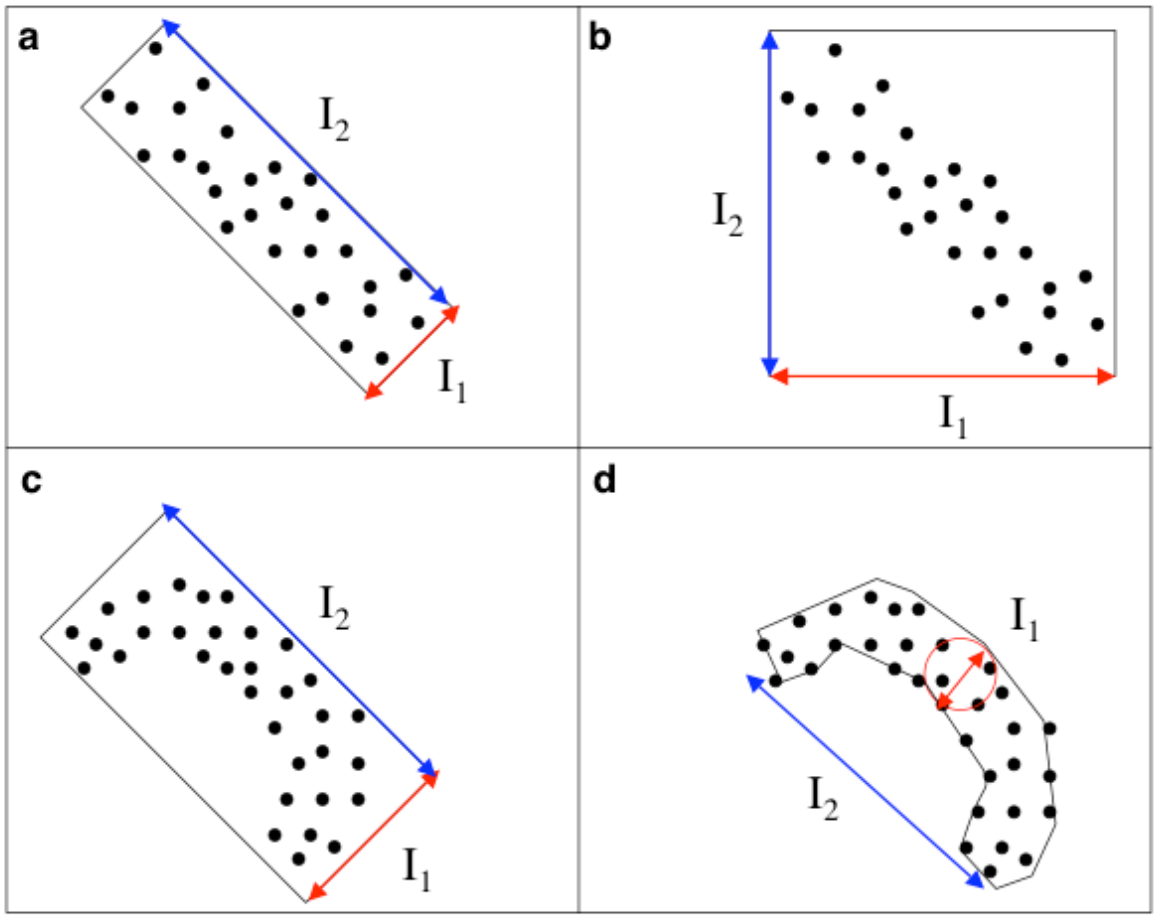

Figure 9. Definition of an aggregation's dimensions
The Minimal Bounding Box (MBB) (a) gives a reliable definition of the dimensions when the aggregation is convex. An arbitrary Cartesian bounding box, on the other hand, is in general completely unreliable (b). For non-convex aggregations the MBB fails to detect the real dimensions. For example, in the figure it overestimates the aggregation's thickness, i.e. the dimension of the shortest axis (c). In order to give an accurate measure of thickness one needs to measure the diameter of the largest sphere entirely contained in the aggregation (d).